\documentclass[final,authoryear,3p,times,twocolumn]{elsarticle}

\usepackage[latin2]{inputenc}
\usepackage{graphicx}

\usepackage{amssymb}
\usepackage{amsthm}
\usepackage{ulem}

\journal{Applied Radiation and Isotopes 115(2016)262}

\begin{document}

\begin{frontmatter}



\title{Activation cross sections of proton and deuteron induced nuclear reactions on holmium and erbium, related to the production of $^{161}$Er and $^{160}$Er medical isotopes}


\author[1]{F. T\'ark\'anyi}
\author[1]{F. Ditr\'oi\corref{*}}
\author[1]{S. Tak\'acs}
\author[2]{A. Hermanne}
\author[3]{M. Baba}

\cortext[*]{Corresponding author: ditroi@atomki.hu}

\address[1]{Institute for Nuclear Research, Hungarian Academy of Sciences (ATOMKI),  Debrecen, Hungary}
\address[2]{Cyclotron Laboratory, Vrije Universiteit Brussel (VUB), Brussels, Belgium}
\address[4]{Institute of Physics and Power Engineering (IPPE), Obninsk, Russia}
\address[3]{Cyclotron Radioisotope Center (CYRIC), Tohoku University, Sendai, Japan}

\begin{abstract}
\noindent Experimental excitation functions for long-lived products in proton induced reactions were measured with the activation method in the 37-65 MeV energy range on natural holmium. Stacked foil irradiation technique and high resolution gamma spectrometry were used in order to measure cross-section data for the production of $^{161}$Er, $^{160}$Er and $^{159,157}$Dy. For comparison of the production routes of medically related $^{161}$Er and $^{160}$Er radioisotopes new experimental cross section data were deduced for the $^{162}$Er(p,x)$^{161,160}$Er and $^{162}$Er(d,x)$^{161,160}$Er reactions by re-evaluating  gamma-ray spectra from earlier measurements. No earlier data were found in the literature for these reactions. The experimental data are compared with results of TALYS theoretical code reported in TENDL-2015.
\end{abstract}


\end{frontmatter}


\section{Introduction}
\label{1}
In connection with research projects on activation cross-sections for different applications, we are performing a systematic experimental study of proton and deuteron induced activation cross-sections on different targets. Today, this study involves several hundreds of reactions taking place on the stable isotopes of more than 50 elements. We made first systematic study of proton and deuteron induced reactions on lanthanides, resulting in among others, isotope production data for $^{177}$Lu, $^{169}$Yb, $^{167}$Tm, $^{171}$Er, $^{165}$Er, $^{160}$Er, $^{166}$Ho, $^{161}$Ho, $^{166}$Dy, $^{159}$Dy, $^{157}$Dy, $^{149}$Tb, $^{152}$Tb, $^{155}$Tb, $^{161}$Tb, $^{153}$Gd, $^{140}$Nd, $^{149}$Pm, $^{147}$Pm, $^{139}$Pr, $^{153}$Sm, $^{145}$Sm, $^{143}$Pr, $^{142}$Pr, $^{141}$Ce, $^{139}$Ce, $^{137m}$Ce and $^{134}$Ce/$^{134}$La.  Concerning nuclear reactions on monoisotopic $^{165}$Ho we performed experiments on alpha particle induced reactions \citep{Tarkanyi2010}, on deuteron induced activation data \citep{Tarkanyi2014,Tarkanyi2008}. Earlier we investigated the activation cross sections of proton induced reactions on holmium up to 36 MeV \citep{Tarkanyi2008c} in connection with the production of medically related $^{165}$Er \citep{Beyer}. In the present study we have extended the energy range  up to 65 MeV in connection with the investigation of the production routes of medically related $^{161}$Er($^{161}$Ho) \citep{Tarkanyi2013} and $^{160}$Er \citep{Rosch}. The production of $^{161}$Er has importance for indirect production of $^{161}$Ho (T$_{1/2}$ = 2.48 h, electrons: 32 keV) therapeutic radioisotope. The $^{160}$Er (T$_{1/2}$ = 28.58 h, electrons: 6.5 keV) was also suggested a candidate for internal radiotherapy (see \citep{Rosch}.
During the evaluation of the production routes of $^{160}$Er we found unpublished data for production of these radioisotopes by $^{nat}$Er(p,x) and $^{nat}$Er(d,x) reactions. The results for other reaction products of  $^{nat}$Er(p,x) and $^{nat}$Er(d,x) reactions have already been published \citep{Tarkanyi2007b, Tarkanyi2008c}.

\section{Experimental}
\label{2}

The irradiation of holmium samples was performed in 2014 at the LLN Cyclone110 cyclotron (Louvain la Neuve, Belgium) with 65 MeV protons, for 1 h at 35 nA beam intensity. The stack contained 19 sets of 10 $\mu$m Al, 116 $\mu$m In, 99.2 $\mu$m Al, 8.41 $\mu$m V, 99.2 $\mu$m Al, 26.2 $\mu$m Ho and 99.2 $\mu$m Al foils. The Al foils served as monitors and for catching of possible recoiled isotopes. The median proton energy in the last Ho foil in the stack was calculated as 36 MeV \citep{Andersen}.
The stacks were mounted in a Faraday cup like target holder equipped with a long collimator.  The activity of all produced radionuclides, both in the target material, and in the Al-backings/degraders/monitor foils were measured without chemical separation by using HPGe detector based spectrometers. 
The measurements started at nearly EOB + 10 h due to high dose rate of the irradiated targets and the transport time from LLN to VUB premises (Vrije Universiteit Brussel, Brussels, Belgium), hence only longer-lived activation products could be identified. Due to the multi-target irradiation, the large number of different target foils and the limited detector capacity, optimization of gamma-spectra measurements for all products was practically impossible.
The data evaluation was performed in a standard way.  Direct and cumulative cross-sections, depending on the contributing processes, were calculated from the activity measurements by using standard activation and decay formulas. The nuclear decay data used are collected in the Table 1 and were taken from \citep{Nudat}. The reaction Q-values were determined by using the NNDC Q-value calculator \citep{Pritychenko}.
The number of incident particles and the energy scale, initially derived from direct current measurement and accelerator settings, were corrected \citep{Tarkanyi1991} from the  re-measured cross-sections over the whole energy range of the $^{nat}$Al(p,x)$^{22,24}$Na monitor reactions compared with the recommended data from IAEA-TECDOC-1211 \citep{Tarkanyi2001}.
The uncertainties of the cross section values was estimated by the standard technique according to the recommendations of the ISO guide \citep{Inter}.
The uncertainty of the energy scale was estimated by taking into account the energy uncertainty of the primary beam, the possible variation in the target thickness and the effect of beam straggling and cross checked through the re-measured excitation function of monitor reaction.
The experimental and data evaluation of the $^{nat}$Er(p,x) reaction are described in detail in \citep{Tarkanyi2008c} and are summarized here. The irradiations were done in 2006 at an external beam of CYRIC AVF 110 (Sendai, Japan) cyclotron at 70 MeV, for 1h at 58 nA beam intensity. The stack contained 13 stacked sets of Er (25 $\mu$m), Co (50 $\mu$m), Al (10 $\mu$m), In (50 $\mu$m) and Al (100 $\mu$m) foils.  The covered energy range for erbium was 69.5-59.4 MeV.  The radioactivity of each sample and monitor foil was measured by HPGe $\gamma$-spectrometers.
The investigation of activation cross sections of $^{nat}$Er(d,x) were done at the external beam lines of the cyclotrons of the VUB (E$_d$ = 21 MeV) and of  CYRIC (E$_d$ = 40 MeV). The radioactivity of each sample and monitor foil was measured non-destructively by HPGe $\gamma$-spectrometers. The covered energy range was 40 -16 MeV. Commercial (Goodfellow) high purity Er foils (25 and 32 $\mu$m) were stacked together with Ti (12 and 31 $\mu$m) and Al (100 $\mu$m) monitor foils. The spectrum measurements and the data evaluation were similar to described above.  More details can be found in \citep{Tarkanyi2007, Tarkanyi2008c}. The used decay data and the reaction Q-values are also collected in Table 1.

\begin{table*}[t]
\tiny
\caption{Decay data of the produced radioisotopes and contributing production reactions \citep{Nudat, Pritychenko}}
\begin{center}
\begin{tabular}{|p{0.3in}|p{0.3in}|p{0.6in}|p{0.8in}|p{0.6in}|p{0.6in}|p{0.7in}|p{0.6in}|p{0.4in}|} \hline 
\textbf{Isotope\newline I$^{\pi}$\newline T$_{1/2}$\newline } & \textbf{Decay\newline \%} & \textbf{E$_\gamma$ (I$_\gamma$)\newline keV(\%)} & \textbf{Contributing reactions\newline $^{165}$Ho(p,x)} & \textbf{Q-value\newline (keV)\newline $^{165}$Ho(p,x)} & \textbf{Contributing reactions\newline $^{nat}$Er(p,x)} & \textbf{Q-value\newline (keV)\newline $^{nat}$Er(p,x)} & \textbf{Contributing reactions\newline ${}^{nat}$Er(d,x)} & \textbf{Q-value\newline (keV)\newline $^{nat}$Er(d,x)} \\ \hline 
$^{161}$Er\newline 3/2$^{-}$\newline 3.21 h & EC: 100 & 211.15\newline 592.6\newline 826.6\newline 931.7 & (p,5n) & -32763.25 & $^{162}$Er(p,pn)\newline ${}^{164}$Er(p,p3n)\newline $^{166}$Er(p,p5n)\newline $^{167}$Er(p,p6n)\newline $^{168}$Er(p,p7n)\newline $^{170}$Er(p,p9n)\newline ${}^{161}$Tm decay & -9204.23\newline -24955.25\newline -40081.87\newline -46518.34\newline -54289.66\newline -67550.67\newline -13288.2 &  &  \\ \hline 
$^{160}$Er\newline 0${}^{+}$\newline 28.58 h & $\varepsilonup$:100 & $^{160g}$Ho $\gamma$-lines\newline 645.40 \newline 646.3 2\newline 728.18 2 \newline 879.383 \newline 962.317\newline 966.171 & (p,6n) & -39973.0 & $^{162}$Er(p,p2n)\newline ${}^{164}$Er(p,p4n)\newline $^{166}$Er(p,p6n)\newline $^{167}$Er(p,p7n)\newline ${}^{168}$Er(p,p8n)\newline $^{160}$Tm decay & -16411.6\newline -32162.6\newline -47289.3\newline -53725.7\newline -61497.0\newline -22956.2 & $^{162}$Er(d,p3n)\newline $^{164}$Er(d,p5n)\newline $^{166}$Er(d,p7n)\newline $^{167}$Er(d,p8n)\newline $^{168}$Er(d,p9n)\newline $^{160}$Tm decay & -18636.2\newline -34387.2\newline -49513.8\newline -55950.3\newline -63721.6\newline -25180.7 \\ \hline 
${}^{157}$Dy\newline 3/2$^{-}$\newline 8.14 h & $\varepsilonup$:100 & 326.336 (~93) & (p,2p7n)\newline ${}^{157}$Ho decay & -59262.85\newline -62638.3 &  &  &  &  \\ \hline 
${}^{159}$Dy\newline 3/2$^{-}$\newline 144.4 d & $\varepsilonup$:100 & 58.0  & (p,2p5n)\newline ${}^{159}$Ho decay & -43377.58\newline -45997.53 &  &  &  &  \\ \hline 
\end{tabular}

\begin{flushleft}
\tiny{\noindent When complex particles are emitted instead of individual protons and neutrons the Q-values have to be decreased by the respective binding energies of the compound particles: np-d: +2.2 MeV; 2np-t: +8.48 MeV; n2p-$^3$He: +7.72 MeV; 2n2p-$\alpha$: +28.30 MeV
}
\end{flushleft}
\end{center}
\end{table*}




\section{Experimental results, excitation functions}
\label{3}

For the experiments on Ho targets most of the activation products are short-lived compared to the rather long cooling time that had to be respected. Therefore, only excitation functions of a few longer-lived (T$_{1/2}$ $>$ 3 h) could be measured. We have tried to deduce cross-sections for production of  the medically relevant $^{165}$Er though the emitted X-rays (already investigated earlier, see \citep{Tarkanyi2008b}, but similar X-rays are emitted from the longer-lived $^{160}$Er, induced because of the higher energy irradiation in this experiment. Separation of the signals from the two isotopes is in principle possible but only with high uncertainties. It should also be mentioned, that from point of view of practical production the cross-section data at high energy for $^{165}$Ho(p,n)$^{165}$Er are not so important, as the maximum of this reaction is below 20 MeV.
The numerical values of the recently measured cross-sections of the investigated reactions are collected in Table 2. The excitation functions are shown in Figs. 1-4The cross sections are compared with the theoretical data available in the TENDL-2014 and TENDL-2015 on-line libraries \citep{Koning2015, Koning2014} based on TALYS 1.6 code \citep{Koning2012}.

\subsection{The $^{165}$Ho(p,5n)$^{161}$Er, $^{162}$Er(p,x)$^{161}$Er and $^{162}$Er(d,x)$^{161}$Er  reactions}
\label{3.1}
No cross section data have been reported earlier for production of the $^{161}$Er (T$_{1/2}$= 3.21 h) (Fig. 1) produced by these reactions. The experimental threshold for the $^{165}$Ho(p,5n) reaction is in good agreement with the calculated one (Fig. 1). Some energy shift can be observed with the predictions in TENDL-2015 for this reaction. According to Fig. 1 by bombarding $^{nat}$Er up to 25 MeV proton energy and 35 MeV deuteron energy nuclear reactions only on $^{162}$Er contribute to the production of $^{161}$Er.  Therefore, from so called elemental cross sections isotopic cross section can be deduced.  The values in TENDL-2015 are in acceptable agreement with the experimental data in the case of deuteron reactions on erbium, but cannot predict the neither the trend nor the values in the case of proton induced reactions. The comparison with the TENDL calculations for other bombarding particles and with the previous deuteron induced experimental data shows that the proton induced production route is a good alternative in the case of high energy proton accelerators, but at lower energy deuterons accelerators the deuteron route is preferable. 

\begin{figure}
\includegraphics[width=0.5\textwidth]{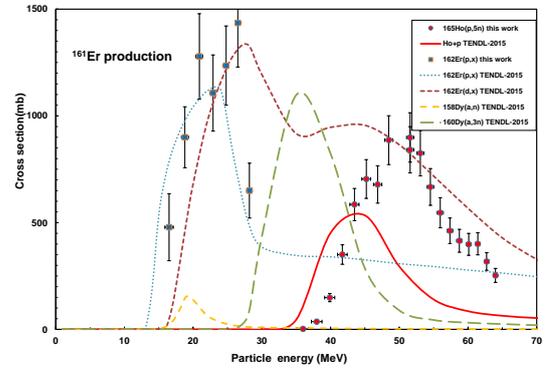}
\caption{Measured excitation function of the $^{165}$Ho(p,5n)$^{161}$Er reaction in comparison with the theory, and with other production routes (TENDL-2015) $^{161}$Er and with our previous experimental results on $^{162}$Er(d,x)}
\label{fig:1}       
\end{figure}

\subsection{The $^{165}$Ho(p,6n)$^{160}$Er, $^{nat}$Er(p,x)$^{160}$Er and $^{nat}$Er(d,x)$^{160}$Er reactions}
\label{3.2}
The amount of $^{160}$Er produced in the proton experiment was determined through the $^{160}$Er(28.58 h) $\longrightarrow$ $^{160m}$Ho(5.02 h) $\longrightarrow$ $^3{160g}$Ho(25.6 min) decay chain. The $^{160}$Er (28.58 h) radioisotope decays for 100 \% by EC to the metastable state of $^{160}$Ho (5.02 h, E$_x$ = 59.983 keV). The quantification through the 59 keV gamma-line was impossible due to the overlapping X-ray lines. The $^{160m}$Ho isomeric state decays with IT: 73 \% to the ground state of $^{160}$Ho (25.6 min). The decay of the $^{160}$Ho ground state is followed by strong high energy gamma-lines, which were used for determination of cross sections of $^{160}$Er, by using spectra measured after the complete decay of directly produced $^{160m}$Ho and $^{160g}$Ho. The measured excitation functions are shown in Fig. 2, in comparison with the other production routes and with the results of the theoretical model code calculations. From Fig. 2 it is seen that the TENDL-2015 curve cannot describe again the excitation function for the proton induced reaction on holmium, at least it contains a large energy shift. The TENDL-2015 approach for $^{162}$Er+p also contains an energy shift while the deuteron induced reaction on the same is predicted with acceptable agreement. In this case the deuteron induced reactions seem to be more preferable again, while the proton reaction on holmium is only recommended at high energy proton accelerators. 

\begin{figure}
\includegraphics[width=0.5\textwidth]{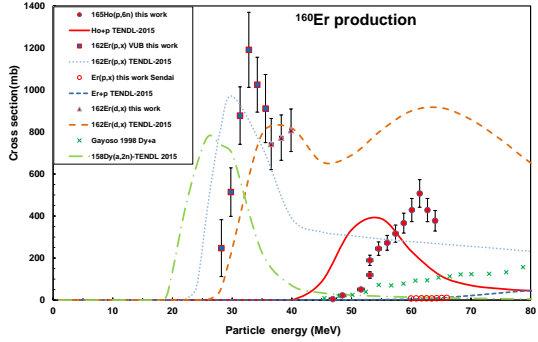}
\caption{Excitation function of the $^{165}$Ho(p,6n)$^{160}$Er reaction in comparison with the theory and with other production routes of $^{160}$Er}
\label{fig:2}       
\end{figure}

\subsection{The $^{165}$Ho(p,2p7n)$^{159}$Dy reaction}
\label{3.3}
The $^{159}$Dy (T$_{1/2}$ = 144.4 d) was identified through its low energy gamma-line (58 keV) with poor statistics. The cumulative cross section includes the decay of the $^{159}$Ho parent (T$_{1/2}$ = 33.05 min) (Fig. 3). There was no previous literature data found for this reaction. The TENDL-2015 curve cannot reproduce the trend of the experimental data, gives acceptable approximation only up to 50 MeV.

\begin{figure}
\includegraphics[width=0.5\textwidth]{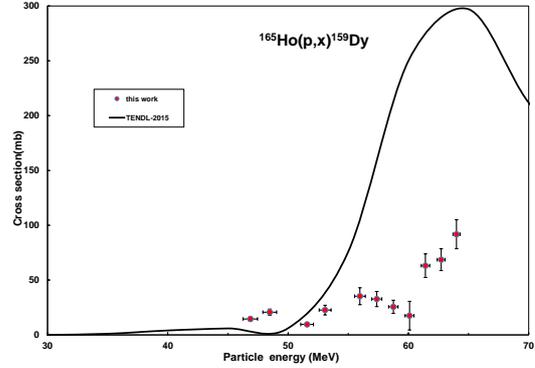}
\caption{Excitation function of the $^{165}$Ho(p,x)$^{159}$Dy reaction in comparison with the theory}
\label{fig:3}       
\end{figure}

\subsection{The $^{165}$Ho(p,2p7n)$^{157}$Dy reaction}
\label{3.4}

In the investigated energy range the $^{157}$Dy (T$_{1/2}$ = 8.14 h) is produced directly and through the decay of $^{157}$Ho (T$_{1/2}$ = 12.6 min). The experimental and theoretical cross sections are shown in Fig. 4. There was no previous literature data found for this reaction. The TENDL-2015 curve cannot reproduce the trend of the experimental data again, gives acceptable approximation only up to 60 MeV.

\begin{figure}
\includegraphics[width=0.5\textwidth]{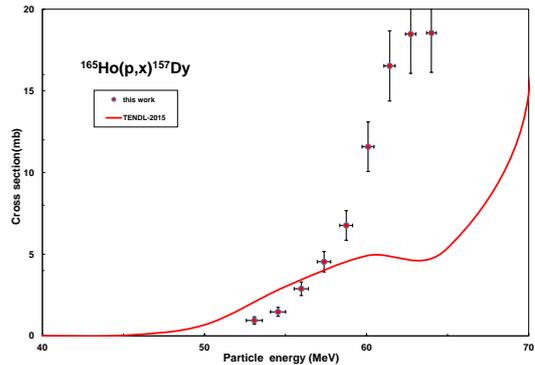}
\caption{Excitation function of the $^{165}$Ho(p,x)$^{157}$Dy reaction in comparison with the theory}
\label{fig:4}       
\end{figure}

\begin{table*}[t]
\tiny
\caption{Experimental cross sections for the $^{165}$Ho(p,5n)$^{161}$Er, $^{165}$Ho(p,6n)$^{160}$Er and $^{165}$Ho(p,x)$^{157}$Dy reactions}
\begin{center}
\begin{tabular}{|p{0.2in}|p{0.2in}|p{0.2in}|p{0.2in}|p{0.2in}|p{0.2in}|p{0.2in}|p{0.2in}|p{0.2in}|p{0.2in}|} \hline 
\multicolumn{2}{|p{0.4in}|}{\textbf{Proton energy\newline E $\pm \Delta$E\newline MeV}} & \multicolumn{8}{|p{1.6in}|}{\textbf{Cross section  $\sigma \pm \Delta\sigma$ \newline mb}} \\ \hline 
\multicolumn{2}{|p{0.4in}|}{\textbf{}} & \multicolumn{2}{|p{0.4in}|}{\textbf{$^{161}$Er}} & \multicolumn{2}{|p{0.4in}|}{\textbf{${}^{160}$Er}} & \multicolumn{2}{|p{0.4in}|}{\textbf{$^{159}$Dy}} & \multicolumn{2}{|p{0.4in}|}{\textbf{$^{157}$Dy}} \\ \hline 
36.0 & 0.8 & 2.5 & 0.3 & ~ &  & ~ &  & ~ &  \\ \hline 
38.0 & 0.8 & 36.3 & 4.7 & ~ &  & ~ &  & ~ &  \\ \hline 
39.9 & 0.7 & 148.9 & 19.3 & ~ &  & ~ &  & ~ &  \\ \hline 
41.7 & 0.7 & 350.5 & 45.4 & ~ &  & ~ &  & ~ &  \\ \hline 
43.5 & 0.7 & 584.9 & 75.4 & ~ &  & ~ &  & ~ &  \\ \hline 
45.2 & 0.6 & 704.5 & 90.9 & ~ &  & ~ &  & ~ &  \\ \hline 
46.9 & 0.6 & 678.6 & 87.5 & 4.6 & 0.6 & 14.5 & 1.9 & ~ &  \\ \hline 
48.5 & 0.6 & 886.6 & 114.3 & 22.2 & 2.9 & 20.7 & 2.8 & ~ &  \\ \hline 
51.6 & 0.5 & 899.0 & 116.0 & 119.3 & 15.5 & ~ &  & ~ &  \\ \hline 
51.6 & 0.5 & 841.3 & 108.5 & 50.8 & 6.7 & 9.6 & 1.3 & ~ &  \\ \hline 
53.1 & 0.5 & 825.3 & 106.5 & 189.5 & 24.5 & 22.7 & 4.4 & 0.9 & 0.2 \\ \hline 
54.5 & 0.5 & 667.3 & 86.1 & 244.9 & 31.7 & ~ &  & 1.5 & 0.3 \\ \hline 
56.0 & 0.4 & 546.4 & 70.6 & 272.7 & 35.3 & 35.3 & 7.7 & 2.9 & 0.4 \\ \hline 
57.4 & 0.4 & 462.0 & 59.7 & 316.5 & 40.9 & 32.7 & 6.9 & 4.5 & 0.6 \\ \hline 
58.7 & 0.4 & 415.3 & 53.7 & 366.6 & 47.4 & 25.6 & 6.0 & 6.8 & 0.9 \\ \hline 
60.1 & 0.4 & 398.4 & 51.5 & 428.4 & 55.3 & 17.6 & 13.1 & 11.6 & 1.5 \\ \hline 
61.4 & 0.3 & 400.7 & 51.8 & 507.5 & 65.5 & 63.1 & 10.8 & 16.5 & 2.2 \\ \hline 
62.7 & 0.3 & 318.0 & 41.2 & 428.3 & 55.3 & 68.6 & 9.9 & 18.5 & 2.4 \\ \hline 
64.0 & 0.3 & 252.9 & 32.8 & 377.4 & 48.7 & 91.8 & 13.2 & 18.5 & 2.4 \\ \hline 
\end{tabular}

\end{center}
\end{table*}

\begin{table*}[t]
\tiny
\caption{Experimental cross sections for the $^{162}$Er(p,x)$^{161}$Er, 162,$^{nat}$Er(p,x)$^{160}$Er and $^{162}$Er(d,x)$^{160}$Er reactions}
\begin{center}
\begin{tabular}{|p{0.2in}|p{0.2in}|p{0.2in}|p{0.2in}|p{0.2in}|p{0.2in}|} \hline 
\multicolumn{2}{|p{0.4in}|}{\textbf{Particle energy\newline E$\pm \Delta$E\newline MeV}} & \multicolumn{4}{|p{0.8in}|}{\textbf{Cross section $\sigma \pm \Delta\sigma$ \newline mb}} \\ \hline 
\multicolumn{6}{|p{1.2in}|}{\textbf{${}^{162}$Er+p}} \\ \hline 
 &  & \multicolumn{2}{|p{0.4in}|}{\textbf{${}^{161}$Er}} & \multicolumn{2}{|p{0.4in}|}{\textbf{${}^{160}$Er}} \\ \hline 
16.5 & 0.6 & 0.7 & 0.2 & ~ &  \\ \hline 
18.8 & 0.5 & 1.3 & 0.2 & ~ &  \\ \hline 
20.9 & 0.5 & 1.8 & 0.3 & ~ &  \\ \hline 
22.9 & 0.4 & 1.6 & 0.3 & ~ &  \\ \hline 
24.8 & 0.4 & 1.7 & 0.3 & ~ &  \\ \hline 
26.5 & 0.4 & 2.0 & 0.3 & ~ &  \\ \hline 
28.2 & 0.4 & 0.9 & 0.2 & 247.1 & 135.4 \\ \hline 
29.8 & 0.3 & 1.1 & 0.2 & 514.0 & 115.4 \\ \hline 
31.3 & 0.3 & 0.5 & 0.2 & 878.2 & 136.8 \\ \hline 
32.8 & 0.3 & 1.5 & 0.3 & 1191.3 & 178.7 \\ \hline 
34.2 & 0.3 & 2.3 & 0.4 & 1025.4 & 130.7 \\ \hline 
35.6 & 0.3 & 5.3 & 0.7 & 911.6 & 162.2 \\ \hline 
\multicolumn{6}{|p{1.2in}|}{\textbf{${}^{nat}$Er+p Sendai}} \\ \hline 
60.0 & 0.5 & ~ &  & 7.7 & 0.9 \\ \hline 
60.8 & 0.5 & ~ &  & 8.3 & 0.9 \\ \hline 
61.7 & 0.5 & ~ &  & 8.6 & 1.0 \\ \hline 
62.6 & 0.5 & ~ &  & 8.8 & 1.0 \\ \hline 
63.4 & 0.4 & ~ &  & 8.2 & 0.9 \\ \hline 
64.3 & 0.4 & ~ &  & 9.2 & 1.0 \\ \hline 
65.1 & 0.4 & ~ &  & 9.9 & 1.1 \\ \hline 
65.9 & 0.4 & ~ &  & 10.9 & 1.2 \\ \hline 
\multicolumn{6}{|p{1.2in}|}{\textbf{${}^{162}$Er+d}} \\ \hline 
36.5 & 0.3 & ~ &  & 742.6 & 122.0 \\ \hline 
38.2 & 0.3 & ~ &  & 773.2 & 107.9 \\ \hline 
39.9 & 0.3 & ~ &  & 808.3 & 101.4 \\ \hline 
\end{tabular}

\end{center}
\end{table*}

\section{Comparison of the production routes of $^{161}$Ho and $^{160}$Er.}
\label{4}

\subsection{$^{161}$Ho}
\label{4.1}

$^{161}$Ho (T$_{1/2}$  = 2.48 h) can be produced directly and indirectly through decay. The direct production routes have already been investigated and reviewed by us \citep{Tarkanyi2013}, where:
\begin{itemize} 
\item	irradiation of $^{159}$Tb by $\alpha$- or $^3$He-particles relying on the $^{159}$Tb($\alpha$,2n)$^{161}$Ho and $^{159}$Tb($^3$He,n)$^{161}$Ho reactions, and
\item	irradiation of dysprosium targets using protons via the $^{161}$Dy(p,n) and $^{162}$Dy(p,2n) reactions  or using deuterons via $^{160}$Dy(d,n) and $^{161}$Dy(d,2n) reactions were discussed.
\end{itemize}
When using the $^{159}$Tb($\alpha$,2n), $^{162}$Dy(p,2n) or $^{161}$Dy(d,2n) processes, also emission of only one neutron can take place, resulting in simultaneous production of $^{162}$Ho, which is a radionuclide impurity. This radionuclide has a short half-life ground state $^{162g}$Ho (T$_{1/2}$  = 15 min) and a longer-lived isomeric state (T$_{1/2}$  = 67.0 min). From the point of view of $^{161}$Ho production, the contamination with the longer-lived excited state is important shortly after EOB, but will be reduced by proper cooling time. The decay through internal transition of $^{162m}$Ho is followed only by low energy, low intensity $\gamma$-ray emission, but the 38 \% electron capture decay results in strong, high energy $\gamma$-lines. 
.
From the excitation functions of the above mentioned reactions the following conclusions can be drawn:
\begin{itemize} 
\item	The production yields for the $^{162}$Dy(p,2n) is the highest followed by the $^{161}$Dy(d,2n), $^{159}$Tb($\alpha$,2n) and $^{161}$Dy(p,n) processes.
\item	No $^{162m}$Ho impurity is produced when using of $^{159}$Tb($^3$He,n), $^{161}$Dy(p,n) and the $^{160}$Dy(d,n) reactions. Among them the $^{161}$Dy(p,n) reaction has the highest cross-section ($\sigma_{max}$ ~ 260 mb) followed by the $^{160}$Dy(d,n) reaction($\sigma_{max}$ ~ 60 mb) and the less productive $^{159}$Tb($^3$He,n) ($\sigma_{max}$ ~ 1 mb).  
\item	The element Tb is monoisotopic, relatively cheap and recovery is practically not necessary. We should add to this review the role of another important radionuclide impurity namely the $^{160m}$Ho.  By one more neutron evaporation in the above mentioned reactions the $^{160}$Ho can be directly produced. The $^{160m}$Ho (T$_{1/2}$  = 5.02 h, 2$^-$, IT 73 \%) $\longrightarrow$ $^{160g}$Ho (T$_{1/2}$  = 25.6 min) decay chain results in high energy $\gamma$-rays. 
\end{itemize}
The second route is indirect: production through the decay of $^{161}$Er. The disadvantage of the indirect production that the half-life of the $^{161}$Er (T$_{1/2}$  = 3.21 h) is relatively short, which limits the irradiation time to be fully exploited as a generator. The production routes of $^{161}$Er at accelerators include proton and deuteron induced reaction on holmium and erbium, $^3$He- and $\alpha$-particle induced reactions on dysprosium.  The comparison of the excitation functions of these production routes is shown in Fig. 1.
Except for the $^{165}$Ho+p route all other reactions require highly enriched targets. As the abundances of $^{162}$Er (0.14 \%), $^{158}$Dy (0.06 \%), $^{160}$Dy (2.34 \%) are very low the highly enriched targets are expensive and target recovery has to be developed and used. From other side the $^{165}$Ho(p,n) reaction requires a high energy accelerator as the practical threshold is over 35 MeV.
For the indirect routes the $^{162}$Ho radionuclide impurity does not exist. But simultaneous production of longer-lived $^{160}$Er results in high energy gamma-lines through the $^{160}$Er (T$_{1/2}$  = 28.58 h)  $\longrightarrow$ $^{160m}$Ho(T$_{1/2}$  = 5.02 h) $\longrightarrow$ $^{160g}$Ho (T$_{1/2}$  = 25.6 min) decay chain.

\subsection{$^{160}$Er}
\label{4.2}
For $^{160}$Er (T$_{1/2}$  = 28.6 h)  the most important production routes are shown in Fig. 2. The $^{162}$Er(p,x) and $^{162}$Er(d,x) reactions have high cross sections, but  the abundance of the $^{162}$Er is very low (0.14 \%). When using low enrichment or natural erbium targets irradiated with deuterons the radionuclide impurities arise from $^{161}$Er (T$_{1/2}$  = 3.24 h), $^{163}$Er (T$_{1/2}$  = 75 min), $^{165}$Er (T$_{1/2}$  = 10.3 h), $^{169}$Er (T$_{1/2}$  = 9.4 d )and $^{171}$Er (T$_{1/2}$  = 7.52 h) are formed. Except the $^{169}$Er isotope all contaminants have shorter half-life than $^{160}$Er, i. e. their amount can be minimized by proper cooling time. The presence of $^{165}$Er and $^{169}$Er in some cases does not disturb very much being therapeutic isotope and has no high energy gammas. Of course in case of erbium targets the product is carrier added.
No carrier added $^{160}$Er can be produced via the $^{165}$Ho(p,6n) reaction at high energy accelerators and alpha induced reactions on dysprosium: $^{158}$Dy($\alpha$,2n) or $^{nat}$Dy($\alpha$,xn).  In case of $^{158}$Dy($\alpha$,2n) there are no long-lived Er by-products. $^{nat}$Dy target can also be used, because the yield is relatively high and the half-life of the simultaneously produced disturbing $^{161}$Er is significantly shorter (T$_{1/2}$ = 3.28 h). The other byproduct, the $^{165}$Er (T$_{1/2}$  = 10.3 h) has no gamma lines and it is also therapeutic radioisotope.

\section{Summary}
\label{5}

The principal aim of this work was an investigation of the production possibility of the radiotherapy related $^{161}$Ho and $^{160}$Er. We present experimental cross-sections for $^{165}$Ho(p,5n)$^{161}$Er, $^{165}$Ho(p,5n)$^{160}$Er, $^{165}$Ho(p,x)$^{159}$Dy  $^{165}$Ho(p,x)$^{157}$Dy  for the first time up to 65 MeV, $^{nat}$Er(p,x)$^{161}$Er  up to 36 MeV, $^{nat}$Er(p,x)$^{160}$Er  up to 70  MeV and  $^{nat}$Er(d,x)$^{160,161}$Er  up to 40 MeV incident particle energies respectively. The TENDL-2015 theoretical data has problems with the prediction the shape of the excitation functions, which can be explained by the general weakness of the code by multiple particle emission and complex particle emission. The energy shifts in Fig. 1 cannot be explained in such a way.
The comparison of the different production routes shows that for production of $^{161}$Ho of high radionuclide purity the $^{161}$Dy(p,n), $^{162}$Dy(p,2n) and $^{161}$Dy(d,2n) reactions give the highest production yields. The $^{162m}$Ho radionuclide impurity level of the last two reactions however is significant. No enriched target material is necessary in case of $^{159}$Tb($\alpha$,2n) (Tb is monoisotopic) but it requires accelerators having medium energy alpha particles. The $^{159}$Tb($^3$He,n) and $^{160}$Dy(d,n) reactions have very low cross sections.
In case of $^{160}$Er the $^{158}$Dy($\alpha$,2n) has higher specific activity and radionuclide purity. In case of $^{165}$Ho(p,6n) the yield is higher, natural target can be used but the specific activity will be lower due to the simultaneous production of the stable $^{162,164}$Er isotopes.

\section{Acknowledgements}
\label{}
This work was done in collaboration of the Institute for Nuclear Research (Debrecen), VUB Brussels and Tohoku University (Sendai). The authors acknowledge the support of the respective institutions in providing the beam time and experimental facilities.
 



\bibliographystyle{elsarticle-harv}
\bibliography{Holmium}







\end{document}